# Secure Quantum Communication: Simulation and Analysis of Quantum Key Distribution Protocols

Mahendra Rasay
*Department of Computer Science and Engineering*
*JAIN (Deemed-to-be) University*
*Bangalore, India*
21btrcs245@jainuniversity.ac.in

Emmanuel D. Sebastian
*Department of Computer Science and Engineering*
*JAIN (Deemed-to-be) University*
*Bangalore, India*
22btrcn086@jainuniversity.ac.in

Subhash Prasad Sah
*Department of Computer Science and Engineering*
*JAIN (Deemed-to-be) University*
*Bangalore, India*
22btrcn284@jainuniversity.ac.in

David Chinamerem Akah
*Department of Computer Science and Engineering*
*JAIN (Deemed-to-be) University*
*Bangalore, India*
22btrcn341@jainuniversity.ac.in

Ajay Kumar Singh*
*Department of Computer Science and Engineering*
*JAIN (Deemed-to-be) University*
*Bangalore, India*
ajay.k.singh@jainuniversity.ac.in

***Abstract*—** *Quantum computing poses significant threats to conventional cryptographic techniques such as RSA and AES, motivating the need for quantum secure communication methods. Quantum Key Distribution (QKD) offers information theoretic security based on fundamental quantum principles. This paper presents a simulation-based analysis of well-known QKD protocols, namely BB84, B92, and E91, using the IBM Qiskit framework. Realistic quantum channel effects, including noise, decoherence, and eavesdropping, are modeled to evaluate protocol performance. Key metrics such as error rate, secret key generation, and security characteristics are analyzed and compared. The study highlights practical challenges in QKD implementation, including hardware limitations and channel losses, and discusses insights toward scalable and robust quantum communication systems. The results support the feasibility of QKD as a promising solution for secure communication in the quantum era.*

**Keywords**: *BB84 protocol, B92 protocol, E91 protocol, Quantum communication, Quantum key distribution, Quantum security, QBER, Quantum simulation*

## I. INTRODUCTION

In the modern digital era, data security has become one of the most critical challenges in communication systems. Every day, vast amounts of sensitive information ranging from personal data to government secrets are transmitted across global networks. These data transmissions rely on cryptographic algorithms such as Rivest–Shamir–Adleman (RSA) and Advanced Encryption Standard (AES) to ensure confidentiality and integrity. However, with the rapid advancement of quantum computing, the security foundations of these classical encryption systems are increasingly being threatened. Quantum computers, leveraging principles such as superposition and entanglement, possess computational capabilities far beyond those of conventional computers. This development poses a major risk to traditional cryptographic methods, as algorithms like RSA, which depend on the difficulty of factoring large prime numbers, can be broken efficiently using Shor's algorithm on a sufficiently powerful quantum computer. Consequently, researchers are now seeking alternative approaches that can maintain secure communication in the era of quantum computing.

Quantum Key Distribution (QKD) has emerged as one of the most promising solutions to this challenge. Unlike classical cryptographic techniques that rely on mathematical complexity, QKD uses the fundamental principles of quantum mechanics to enable two parties to share a secret key securely. The security of QKD is derived from the laws of physics rather than computational hardness. Any attempt to eavesdrop on a quantum communication channel introduces detectable disturbances, allowing legitimate users to identify and eliminate compromised keys. This fundamental difference makes QKD theoretically unbreakable and a cornerstone for the development of quantum safe communication systems.

The most well-known QKD protocols, such as BB84, B92, and E91, have been extensively studied both theoretically and experimentally. Each of these protocols uses distinct quantum properties and encoding methods to establish secure communication. The BB84 protocol, proposed by Bennett and Brassard in 1984, employs two sets of polarization bases for photon transmission, while the B92 protocol, introduced in 1992, simplifies the process by using only two non- orthogonal states. The E91 protocol, proposed by Ekert in 1991, is based on the principle of quantum entanglement, ensuring security through the violation of Bell's inequalities

To address these challenges, the current research project titled "Secure Quantum Communication: Simulation and Analysis of Quantum Key Distribution Protocols" focuses on simulating and analyzing various QKD protocols using modern quantum computing platforms such as IBM Qiskit. These simulation tools allow for the creation of virtual quantum networks that mimic real world conditions, including quantum noise, signal loss, and eavesdropping attempts. By leveraging simulation rather than relying on current quantum hardware which remains limited in terms of qubit count, stability, and error rates this study enables an in-depth evaluation of QKD behavior under diverse operational scenarios. The results provide insights into key parameters such as secret key generation rate, quantum bit error rate (QBER), and overall protocol efficiency.

Furthermore, this research explores critical implementation challenges in developing scalable and practical QKD systems. Hardware imperfections, photon losses in optical fibers, synchronization issues, and the integration of QKD with existing classical communication infrastructures all pose significant barriers to widespread deployment. Through simulation and analysis, the study aims to understand how these factors impact system performance and to propose possible solutions for mitigating their effects. Additionally, the research includes a comprehensive literature review and gap analysis, identifying key areas for future exploration particularly the development of QKD systems that are more robust against environmental noise and adaptable for large scale communication networks.

Ultimately, this project aims to bridge the gap between theoretical research and practical application in quantum communication. By simulating QKD protocols under realistic network conditions, it provides valuable insights into the transition from idealized models to real world systems. The findings not only contribute to a deeper understanding of quantum communication mechanisms but also support the foundation for future quantum secure networks capable of withstanding the computational threats posed by quantum computers. As the world moves closer to the quantum era, research such as this becomes increasingly vital for safeguarding digital communication, ensuring privacy, and enabling the secure exchange of information in a post quantum world.

## II. LITERATURE REVIEW

Early work in QKD was established by Bennett and Brassard through the BB84 protocol, which demonstrated that secure key exchange can be achieved using non orthogonal quantum states and basis incompatibility [1]. This foundational result showed that any eavesdropping attempt introduces detectable disturbances, forming the basis of most subsequent QKD research. Ekert later proposed the E91 protocol, introducing entanglement based QKD in which security is verified through Bell inequality violations rather than error rates alone [2]. Bennett further simplified prepare and measure QKD through the B92 protocol, which uses only two non-orthogonal states, reducing implementation complexity at the cost of lower efficiency [3].

Simulation based studies have played a central role in evaluating the practical performance of these protocols. Numerous works have modeled BB84 under ideal and noisy channels, consistently showing that secure key generation is possible as long as the QBER remains below established thresholds [4,5]. MATLAB and Python based simulations demonstrate how depolarization noise, photon loss, and intercept resend attacks affect BB84 performance, confirming its robustness under moderate noise conditions [6,7]. Extensions of these simulations to hybrid fiber free space optical channels reveal that atmospheric effects such as fog, rain, and beam divergence significantly degrade polarization fidelity and reduce achievable key rates [8].

Entanglement based simulations of the E91 protocol emphasize the role of Bell inequality violation as a security metric. Studies show that while E91 offers stronger theoretical security guarantees, its performance is highly sensitive to detector inefficiencies and channel noise, often resulting in lower key rates compared to BB84 under equivalent conditions [9,10].

Nevertheless, E91 simulations highlight its relevance for device independent and networked QKD scenarios, where trust in measurement devices cannot be assumed [11].

The B92 protocol has been investigated primarily in simplified simulation environments. Results consistently indicate that the use of non-orthogonal states leads to a high fraction of inconclusive measurements, making B92 more sensitive to noise and loss than BB84 [12]. However, its reduced state space and simpler encoding make it attractive for low complexity and educational implementations [13].

Comparative simulation studies consistently rank BB84 as the most efficient and practical protocol among the three, balancing security, robustness, and implementation feasibility [14,15]. E91 is shown to provide deeper security insight through entanglement verification but at increased system complexity, while B92 trades efficiency for simplicity [16].

Recent literature has expanded simulation models to include advanced attack strategies. Machine learning assisted eavesdropping attacks demonstrate that partial information can be extracted with minimal increase in QBER, challenging traditional security assumptions used in BB84 simulations [17]. These results suggest that classical intercept resend models may underestimate adversarial capabilities and that more conservative security margins are required.

Beyond point to point links, several studies investigate QKD in networked environments. Simulations of QKD over complex network topologies reveal scalability limitations due to loss accumulation and key management constraints, reinforcing the need for hybrid architectures combining discrete variable protocols such as BB84 and E91 with emerging approaches [18].

Measurement device independent QKD simulations further address detector side channel vulnerabilities while preserving the conceptual foundations of BB84 and entanglement-based schemes [19].

Overall, the reviewed literature demonstrates that simulation remains an essential tool for analyzing BB84, E91, and B92 under realistic conditions. While theoretical security proofs establish feasibility, simulation studies reveal critical tradeoffs among efficiency, robustness, and implementation complexity that directly influence practical deployment [20].

## III. METHODOLOGY

This work adopts a simulation-based methodology to analyze and compare major QKD protocols under realistic communication conditions. Due to the limitations of current quantum hardware, software based quantum simulation provides a flexible and controlled environment to study protocol behavior, security, and performance. The methodology combines theoretical protocol modeling, quantum circuit simulation, and performance evaluation using standard metrics.

### A. Simulation Environment

The simulations are implemented using IBM Qiskit, a widely used quantum computing framework that supports quantum circuit construction, execution, and measurement. Qiskit enables accurate modeling of quantum states, measurement bases, and noise effects, making it suitable for evaluating QKD protocols without reliance on physical quantum devices. Classical post-processing steps such as basis comparison, sifting, and error estimation are incorporated to closely resemble real-world QKD workflows.

### B. Quantum States and Encoding Bases

Quantum key distribution relies on encoding classical information into quantum states using different measurement bases. In this study, two polarization bases are used, as commonly adopted in BB84 and B92 based protocols.

*1)   BB84 Protocol Encoding Bases:*

*The Z-basis uses* horizontal and vertical photon polarizations to encode bits:

- Bit 0 → |0⟩ → Horizontal polarization (→)    (1)
- Bit 1 → |1⟩ → Vertical polarization (↑)    (2)

This basis corresponds to the standard computational basis in quantum computing:

|0⟩, |1⟩

*The X-basis* uses diagonal polarization angles (±45°) to encode bits:

- Bit 0 → |+⟩ → +45° polarization    (3)
- Bit 1 → |−⟩ → −45° polarization    (4)

These states are superpositions of |0⟩ and |1⟩. These states can be visualized using the Bloch sphere representation, which provides an intuitive geometric interpretation of qubit states and their relative phases.

*2)   B92 Protocol Encoding Bases:*

In quantum mechanics, two quantum states are said to be orthogonal if they can be perfectly distinguished by an appropriate measurement, which occurs when their inner product is zero. Conversely, non-orthogonal states have a non-zero inner product and therefore cannot be perfectly distinguished. In the B92 protocol, Sender encodes classical information using two non-orthogonal polarization states. In a common polarization-based implementation, the encoding is defined as

$$Bit\ 0\ \rightarrow\ |H\rangle\ (0°)\ ,Bit\ 1\ \rightarrow\ |+\rangle\ (45°)$$

Where, |H⟩ denotes horizontal polarization at 0° as shown in Figure 1 and |+⟩ denotes polarization at 45° as shown in Figure 3 These two states are non-orthogonal, satisfying ⟨H | +⟩ ≠ 0.

Sender performs measurements using states orthogonal to Sender's signal states. Specifically, he randomly chooses between the measurement states

$$|V\rangle\ (90°)\ ,|-\rangle\ (-45°)$$

where |V⟩, it is represented by Figure 2 is orthogonal to |H⟩ and |−⟩, it is represented by Figure 4 is orthogonal to |+⟩. When

Receiver detects a photon in one of these orthogonal states, he can unambiguously infer which state Sender did not send, and hence determine the transmitted bit value. Measurement outcomes that do not yield a conclusive result are discarded.

*3)    E91 Protocol Encoding Bases:*

To maximize Bell-inequality violation, Sender and Receiver carefully chose polarization angles:

Sender's Bases :
- A0: 0° (Key generation basis)
- A1: 0°
- A2: 45°

Receiver's Bases :
- B0: 0° (Key generation basis)
- B1: 22.5°
- B3: 67.5°

These angles are known to produce maximal CHSH violation for Bell states such as $\Phi^+$.

Qubit 0

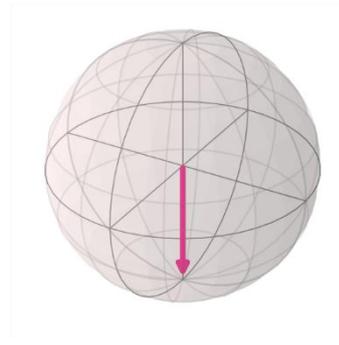

*Fig. 2. Bloch sphere representation of the computational basis state $|1\rangle$*

Figure 2 illustrates the $|1\rangle$ state, which is represented by a vector pointing toward the south pole of the Bloch sphere. This corresponds to $\theta = \pi$, indicating alignment with the $-Z$ direction. As describe by (2) a computational basis state $|1\rangle$ yields the measurement result 1 with certainty when measured in the Z basis.

Qubit 0

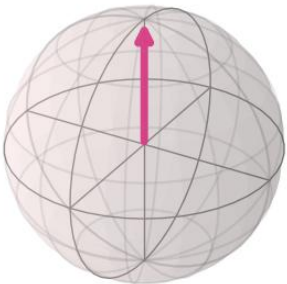

*Fig. 1. Bloch sphere representation of the computational basis state $|0\rangle$*

Figure 1 shows the qubit in the $|0\rangle$ state, represented as a vector pointing toward the north pole of the Bloch sphere. The state $|0\rangle$ corresponds to the polar angle $\theta = 0$, indicating full alignment with the +Z axis. As describe by (1) no superposition or phase is present, and measurement in the Z basis always produces the outcome 0.

Qubit 0

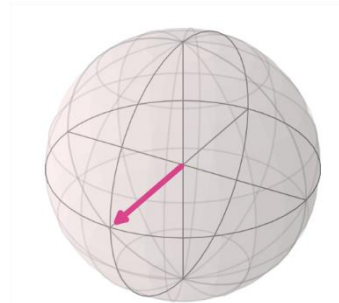

*Fig. 3. Bloch sphere representation of the diagonal basis state $|+\rangle$*

Figure 3 shows the qubit in the diagonal basis state $|+\rangle = \frac{|0\rangle + |1\rangle}{\sqrt{2}}$, represented by a vector lying on the equator and pointing along the +X axis. As describe by (3) the equal amplitudes of $|0\rangle$ and $|1\rangle$ indicate that $|+\rangle$ is an equal superposition with zero relative phase. Measurement in the diagonal (X) basis yields + with probability 1.

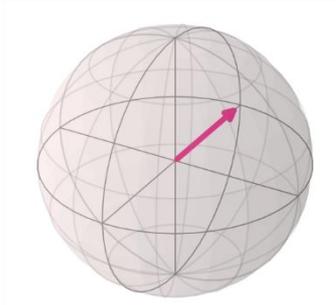

Qubit 0

*Fig. 4. Bloch sphere representation of the diagonal-basis state |−⟩|*

Figure 4 depicts the state $|-\rangle = \frac{|0\rangle - |1\rangle}{\sqrt{2}}$, represented by a vector on the equator pointing along the −X axis. As describe by (4) this position reflects equal probabilities for |0⟩ and |1⟩, but with a π relative phase between them. Measurement in the diagonal (X) basis returns the outcome with certainty.

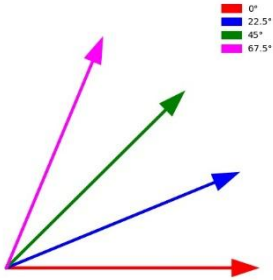

*Fig. 5. Two-dimensional representation of polarization measurement angles*

Figure 5 presents a two-dimensional geometric representation of the polarization measurement bases employed by the Sender and Receiver in the E91 protocol. Each measurement direction is indicated by a color-coded arrow, where the red, blue, green, and magenta arrows correspond to polarization angles of 0°, 22.5°, 45°, and 67.5°, respectively, measured with respect to a common reference axis. This visualization highlights the relative angular separations among the selected measurement bases used for key generation and Bell-test rounds. The intentional choice of non-orthogonal angles is essential for achieving maximal violation of the Bell–CHSH inequality when measurements are performed on maximally entangled Bell states such as $\Phi^+$.

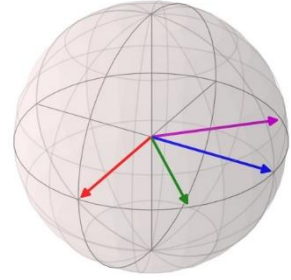

*Fig. 6. Three-dimensional Bloch sphere representation of measurement directions*

Figure 6 presents a three-dimensional Bloch sphere representation of the polarization measurement directions employed by the Sender and Receiver in the E91 quantum key distribution protocol. The measurement bases are depicted as color-coded vectors on the surface of the Bloch sphere, where the red, blue, green, and magenta arrows correspond to polarization angles of 0°, 22.5°, 45°, and 67.5°, respectively. This representation provides a quantum-state interpretation of the measurement settings, illustrating how each polarization basis corresponds to a distinct projective measurement on the qubit state. The relative orientations of the measurement vectors on the Bloch sphere highlight the non-orthogonal nature of the selected bases, which is crucial for achieving maximal violation of the Bell CHSH inequality when measuring maximally entangled Bell states such as $\Phi^+$.

C. Quantum Key Distribution Protocols

*1) BB84 Protocol:*

The BB84 Quantum Key Distribution (QKD) protocol, introduced in 1984 by Charles H. Bennett and Gilles Brassard, represents the first practical application of quantum mechanics to cryptography. Unlike classical cryptographic systems whose security relies on computational hardness assumptions, BB84 offers information theoretic security, guaranteed by the fundamental laws of quantum physics.

In the BB84 protocol, two legitimate parties commonly referred to as Subhash (sender) and David (receiver) establish a shared secret key by transmitting quantum states over a quantum channel and performing classical post-processing over a public but authenticated classical channel. Any attempt by an adversary (Eve) to intercept or measure the quantum states inevitably introduces disturbances that can be detected by the communicating parties.

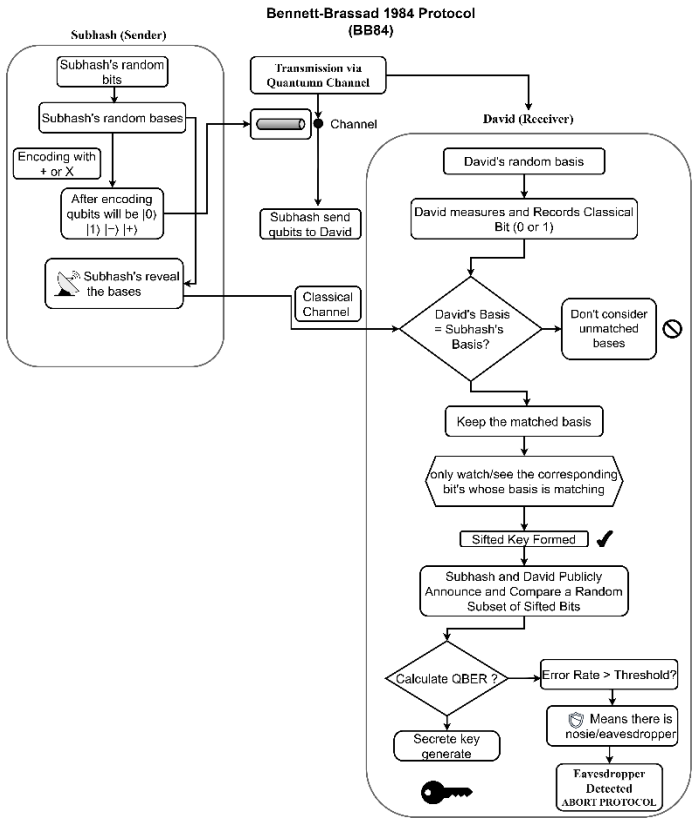

*Fig. 7. BB84 QKD Protocol Workflow.*

TABLE I. *Sifted Key Generation in BB84 Protocol*

| | 1 | 0 | 0 | 1 | 1 | 0 | 1 | 1 |
|---|---|---|---|---|---|---|---|---|
| Sender's random encoding bases | + | × | × | + | + | + | × | × |
| Qubits after encoding | $|1\rangle$ | $|+\rangle$ | $|+\rangle$ | $|1\rangle$ | $|1\rangle$ | $|0\rangle$ | $|-\rangle$ | $|-\rangle$ |
| Sender's polarization | ↑ | ↗ | ↗ | ↑ | ↑ | → | ↘ | ↘ |
| Receiver's random Measurement basis | + | + | × | × | + | + | × | + |
| Receiver's polarization | ↑ | ↘ | ↗ | ↑ | → | → | ↘ | ↗ |
| Sifted Key | 1 | 1 | 0 | 1 | 1 | 0 | 1 | 0 |

*Working of the BB84 Protocol:*

- *Subhash's random bit-string:*
  This is the initial sequence of classical bits chosen randomly by Subhash.
  Subhash's string in the table is:
  1 0 0 1 1 0 1 1

- *Subhash's random encoding bases:*
  For each bit, Subhash randomly selects either the rectilinear (+) basis or the diagonal (×) basis.
  The choice of basis determines how the classical bit is converted into a quantum state.

- *Qubits after encoding:*
  Using the chosen basis, each classical bit is mapped to a quantum state:
  In the rectilinear basis, $0 \to |0\rangle$ and $1 \to |1\rangle$
  In the diagonal basis, $0 \to |+\rangle$ and $1 \to |-\rangle$
  This column shows the resulting encoded states such as $|1\rangle, |+\rangle, |-\rangle$, etc.

- *Subhash's polarization:*
  Each qubit state corresponds to a physical photon polarization. For example:
  $|0\rangle \to 0°$
  $|1\rangle \to 90°$
  $|+\rangle \to 45°$
  $|-\rangle \to 135°$

- *David's random measurement basis:*
  David independently and randomly chooses a basis (+ or ×) to measure each photon.
  Only when David's basis matches Subhash's basis does he obtain the correct bit.

This figure 7 workflow diagram illustrates the BB84 quantum key distribution protocol between a sender (Subhash) and a receiver (David). Subhash generates random bits and bases, encodes qubits using either the standard (Z) or diagonal (X) basis, and transmits these qubits to David via a quantum channel. David independently selects random bases to measure the received qubits and records classical bits accordingly. After the transmission, Subhash reveals the bases used through a classical channel. David and Subhash then compare their respective bases and keep only those bits where their bases matched, discarding unmatched ones. Any subset of these sifted bits is disclosed and compared to check for eavesdropping by calculating the quantum bit error rate (QBER).

*a) Sifted Key Generation Analysis on BB84:* In QKD systems, not all transmitted bits contribute to the final secret key. Only those bits for which the sender and receiver select compatible bases are retained, forming the *sifted key*. To demonstrate this process, a representative BB84 simulation instance is analyzed.

The Table I. Sifted Key Generation includes Subhash's random bit-string, his randomly chosen encoding bases, the resulting qubit states after encoding, and the corresponding photon polarizations. David's measurement bases and detected polarizations are also listed, followed by the final sifted key obtained after comparing bases.

- *David's polarization (measurement outcomes)*
  The observed photon polarizations are listed. If David measures in the wrong basis, the outcome is random.

- *Sifted Key*
  The final row shows the resulting key bits after comparing bases and discarding mismatches.
  The generated sifted key in the table is:
  1 1 0 1 1 0 1 0

*b) Quantum Bit Error Rate (QBER):* The Quantum Bit Error Rate (QBER) quantifies the fraction of mismatched bits between Subhash and David in the sifted key. It serves as a critical indicator of channel noise and potential eavesdropping.

Mathematical Expression

$$QBER = \frac{N_{error}}{N_{total}} \times 100\%$$

Security Threshold

For the BB84 protocol:

- QBER ≤ 11% → Secure key distillation possible
- QBER > 11% → Communication not secure

This threshold arises from rigorous security proofs and guarantees unconditional secrecy after error correction and privacy amplification.

*2) E91 Protocol:*

The E91 protocol is an entanglement-based Quantum Key Distribution (QKD) scheme proposed by Artur Ekert in 1991. Unlike prepare-and-measure protocols, E91 derives its security from quantum entanglement and the violation of Bell inequalities, rather than from disturbance alone. A source generates pairs of entangled photons and distributes one photon to the sender and the other to the receiver. Both parties independently choose random measurement bases and record their outcomes. Using a public classical channel, they reveal only their measurement bases and retain the results corresponding to compatible settings. Security is verified through Bell inequality testing, which allows the detection of any eavesdropping attempt. If a Bell inequality violation is observed, the correlated outcomes are used to generate a secure key, otherwise the protocol is aborted.

In this work, an E91 protocol simulation is implemented using entangled photon pairs, with security evaluated through:

- The Bell–CHSH parameter (S), and
- The Quantum Bit Error Rate (QBER)

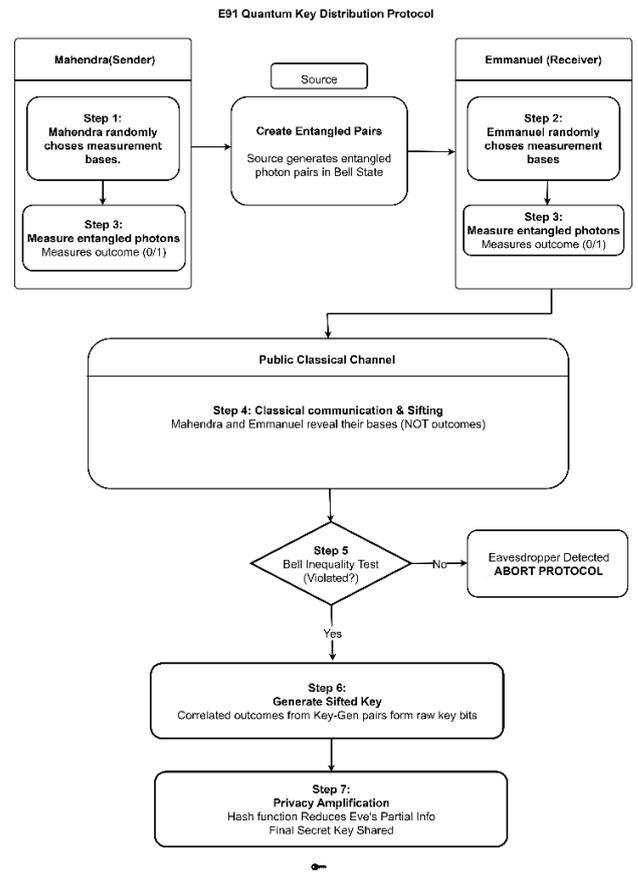

*Fig. 8. E91 QKD Protocol Workflow*

Figure 8 workflow outlines the E91 quantum key distribution protocol, where a source creates pairs of entangled photons shared between the sender (Mahendra) and receiver (Emmanuel). Both Sender and Receiver independently choose random measurement bases and measure their respective entangled photons, recording outcomes as classical bits (0 or 1). Through a classical public channel, Sender and Receiver reveal only their measurement bases, not the measurement results, and sift the data to keep bits where their bases matched. They perform a Bell inequality test to detect any violations, which would indicate the presence of an eavesdropper. If no violation is found, they generate a sifted key from the correlated photon measurements and apply privacy amplification to further secure the final shared secret key. If the Bell is violated, the protocol is aborted to maintain security.

*a) Bell–CHSH Inequality as a Security Test:* The Clauser–Horne–Shimony–Holt (CHSH) inequality is a quantitative test that distinguishes classical (local realistic) correlations from quantum entanglement.

The CHSH inequality is a mathematical test used to check whether correlations between two distant systems can be explained by classical physics (local hidden variables) or require quantum mechanics.

Classical world: correlations are limited

Quantum world: entanglement allows stronger correlations

The CHSH value is:

$S=E(A,B)+(A,B')+E(A',B)-E(A',B')$

where:

E (Ai, Bj) denotes the expectation value of correlated measurement outcomes obtained using measurement settings $A_i$ and $B_j$.

$$E(A_i, B_j) = \frac{N_{same} - N_{diff}}{N_{total}}$$

$N_{Same}$ = number of rounds where sender and receiver got the same outcomes

$N_{Diff}$ = number of rounds where sender and receiver got the different outcomes

$N_{Total} = N_{Same} + N_{Diff}$

*b) Classical and Quantum Bounds:*

Classical (local hidden variables):   |S|≤2
Quantum mechanics:           |S|≤2√2≈2.828

A measured value of S≥2 confirms non-classical correlations and certifies the presence of quantum entanglement.

*c) Entangle Photon Source in E91:* A defining feature of the E91 protocol is the presence of a quantum source that generates entangled photon pairs.

*d) Bell States as Maximally Entangled States:* For a two-qubit system, there exist four maximally entangled states, known as the Bell states. These states form an orthonormal basis for the two-qubit Hilbert space and represent the strongest possible quantum correlations allowed by quantum mechanics.
The four Bell states are defined as:

$$|\Phi^+\rangle = \frac{1}{\sqrt{2}}(|00\rangle + |11\rangle)$$

$$|\Phi^-\rangle = \frac{1}{\sqrt{2}}(|00\rangle - |11\rangle)$$

$$|\varphi^+\rangle = \frac{1}{\sqrt{2}}(|01\rangle + |10\rangle)$$

$$|\varphi^-\rangle = \frac{1}{\sqrt{2}}(|01\rangle - |10\rangle)$$

Each Bell state exhibits perfect quantum entanglement but differs in the correlation pattern between the measurement outcomes of the two particles. Choice of the $\Phi^+$ Bell State in this Simulation. The $\Phi^+$ Bell state is used as the entangled resource for the E91 protocol simulation.

*e) Bell Ratio :* Bell ratio is the fraction of entangled photon pairs reserved only for Bell-inequality testing, not for key generation.

The purpose of Bell test:

The Bell-test fraction controls how many entangled pairs are sacrificed to verify nonlocal correlations via CHSH increasing this fraction strengthens security at the cost of key rate.

In E91:

- We do not use all photons for the key
- Some photons are sacrificed to test security
- This is essential for detecting Eve

Without Bell-test rounds:

- Eve could send classical correlated signals
- Alice & Bob might still see low QBER
- Security would be illusionary

With Bell-test rounds:

- Alice & Bob check nonlocal correlations
- Eve cannot fake Bell violation
- Security is guaranteed by physics

| Bell Ratio | Interpretation |
|---|---|
| 0.1 | Minimal security test |
| 0.25 | Balanced and Realistic |
| 0.5 | Very strong security, short key |

*f) Measurement Settings in the Simulation:* To maximize Bell-inequality violation, Sender and Receiver carefully chose polarization angles:

*Sender's Bases*
$A0$: 0° (Key generation basis)
$A1$: 0°
$A2$: 45°

*Receiver's Bases*

$B0$: 0° (Key generation basis)
$B1$: 22.5°
$B3$: 67.5°

These angles are known to produce maximal CHSH violation for Bell states such as $\Phi^+$.

*g) Working of the E91 Protocol:*

### Step 0: Classical Pre-Agreement
Before quantum transmission begins, Sender and Receiver publicly agree on:
- Allowed measurement bases
- Basis pairs used for key generation
- Basis pairs used for Bell-test (security) evaluation

This classification is fixed and does not depend on observed results.

### Step 1: Entangled State Distribution
A quantum source generates entangled photon pairs and distributes:
- One photon to Sender
- One photon to Receiver

No classical bits exist at this stage.

### Step 2: Measurement
Sender and Receiver independently:
- Randomly select a measurement basis
- Measure their photon

Record the basis and outcome (+1 or −1)

### Step 3: Public Basis Announcement
Over a classical channel:
- Only measurement bases are announced
- Measurement outcomes remain secret

### Step 4: Data Classification
Each measurement round is classified as:
- Key-generation round
- Bell-test round
- Discarded

This classification follows the pre-agreed rule, not adaptive decision-making.

*Role of the Eavesdropper Model (eve mode)*

To analyze security under different attack strategies, the simulation introduces a configurable eavesdropper (Eve) through the parameter eve mode.

*Eve Attack Modes*

| Eve Mode | Description |
|---|---|
| Key | Eve attacks only key-generation rounds |
| Bell | Eve attacks only Bell-test rounds |
| Both | Eve attacks both key and Bell rounds |

This design allows controlled experimentation on how Eve's behavior impacts QBER and CHSH violation independently.

*Impact on CHSH Value (S)*

Bell-test rounds are exclusively responsible for calculating the CHSH parameter $S$, If Eve attacks Bell rounds
(eve mode = "bell" or "both"):
- Entanglement is degraded
- Correlations weaken
- $S$ decreases and eventually falls below the classical bound

Eavesdropper present $\Rightarrow |S| \leq 2$

When no Bell violation is observed, security cannot be certified, and the protocol is aborted.

*Impact on Quantum Bit Error Rate (QBER)*

QBER is computed from key-generation rounds only:

$$\text{QBER} = \frac{\text{Number of mismatched key bits}}{\text{Total key bits}}$$

- Eve attacking key rounds increases QBER
- Noise and imperfections also increase QBER
- QBER reduces the amount of extractable secure key

However, QBER alone does not certify security in E91.
Complementary Roles of S and QBER

| Parameter | Function |
|---|---|
| CHSH value (S) | Certifies quantum security |
| QBER | Determines key quality and rate |

In E91, Bell violation decides whether a key may be generated QBER decides how much secure key can be extracted this separation is a defining feature of entanglement-based QKD.

*Step 5: Interpretation of Eve and Noise*

Eavesdropper present ⇒ Bell inequality is never violated

Environmental noise ⇒ Bell violation may or may not survive

Therefore E91 is conservative, security is accepted only when non-classical correlations are experimentally verified.

*Step 6: Final Security Decision Rule*

The protocol accepts a key if and only if:

|S|>2 (Bell violation confirmed), and QBER ≤ acceptable threshold Otherwise, the protocol is aborted. Let explore it's with example:

TABLE II. *E91 basis measurement*

| Round | Sender basis | Receiver basis | Sender | Receiver | Eve | Purpose |
|---|---|---|---|---|---|---|
| 1 | A1 | B1 | 0 | 0 | | Key |
| 2 | A1 | B1 | 0 | 1 | ■ | Key |
| 3 | A1 | B3 | 1 | 0 | | Bell |
| 4 | A1 | B3 | 1 | 0 | | Bell |
| 5 | A2 | B1 | 0 | 0 | | Bell |
| 6 | A2 | B1 | 0 | 1 | | Bell |
| 7 | A2 | B3 | 0 | 0 | ■ | Bell |
| 8 | A2 | B3 | 1 | 0 | | Bell |
| 9 | A2 | B3 | 0 | 1 | | Bell |
| 10 | A1 | B1 | 0 | 0 | | Key |

Key rounds (A1, B1)
    Round 1,2,10

Bell-test rounds
    (A1, B3): rounds 3,4
    (A2, B1): rounds 5,6
    A2, B3): rounds 7,8,9

Compute correlations: E (Ai, Bj)
E (A1, B1)
Rounds: 1, 2, 10

| Round | Bits | Outcomes |
|---|---|---|
| 1 | (0,0) | Same |
| 2 | (0,1) | Different |
| 10 | (0,0) | Same |

$$E(A1, B1) = \frac{2-1}{3} = \frac{1}{3} \approx 0.33$$

E (A1, B3)
Rounds: 3, 4

| Round | Bits | Outcomes |
|---|---|---|
| 3 | (1,0) | Different |
| 4 | (1,0) | Different |

$$E(A1, B3) = \frac{0-2}{2} = -1$$

E (A2, B1)
Rounds: 5, 6

| Round | Bits | Outcomes |
|---|---|---|
| 5 | (0,0) | Same |
| 6 | (0,1) | Different |

$$E(A2, B1) = \frac{1-1}{2} = 0$$

E (A2, B3)
Rounds: 7,8,9

| Round | Bits | Outcomes |
|---|---|---|
| 7 | (0,0) | Same |
| 8 | (1,0) | Different |
| 9 | (0,1) | Different |

$$E(A1, B1) = \frac{1-2}{3} = \frac{1}{3} \approx -0.33$$

Compute CHSH Value S:

$$S = E(A1, B1) - E(A1, B3) + E(A2, B1) + E(A2, B3)$$
$$S = 0.33 - (-1) + 0 + (-0.33)$$
$$S = 1$$

Here:
S= 1≤2
No Bell inequality Violation
Protocol must abort

Compute QBER:
$$QBER = \frac{1}{3} \times 100\% = 33.33\%$$

Final interpretation:
- No Bell Violation
- High QBER
- E91 Protocol abort

### 3) B92 Protocol:

The B92 protocol is a quantum key distribution (QKD) protocol proposed in 1992 by Charles H. Bennett. It is a simplified variant of the BB84 protocol and employs only two non-orthogonal quantum states for encoding classical information, in contrast to BB84, which uses four states belonging to two mutually unbiased bases.

Unlike BB84, whose security relies on basis mismatch, the B92 protocol derives its security solely from the fundamental indistinguishability of non-orthogonal quantum states, a direct consequence of the laws of quantum mechanics.

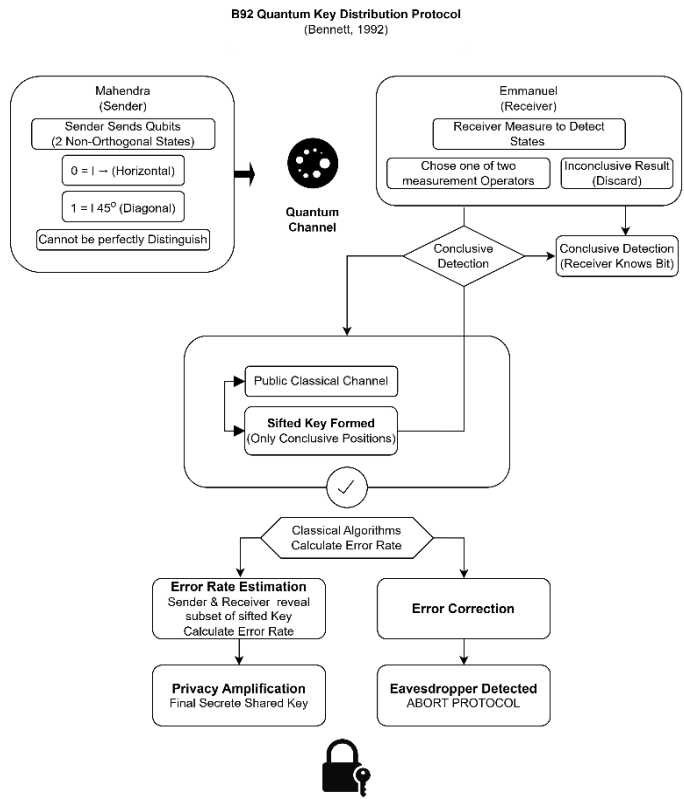

*Fig. 9. B92 QKD Protocol Workflow*

Figure 9 workflow describes the B92 quantum key distribution protocol where the sender (Mahendra) encodes bits using two non-orthogonal quantum states horizontal (0°) and diagonal (45°), corresponding to binary 0 and 1, respectively. The receiver (Emmanuel) measures each qubit using one of two measurement operators, and registers a bit only if the result is conclusive or inconclusive measurement results are discarded. Through a public classical channel, both parties form a "Sifted key" consisting solely of the conclusively detected bits. They then reveal a subset of this key to estimate the error rate if the error rate is acceptable, error correction and privacy amplification steps are applied to generate the final secret key. If the error rate is too high, it signals possible eavesdropping.

*a) Orthogonality and Non-Orthogonality in Quantum Mechanics:* In quantum mechanics, two quantum states $|\psi\rangle$ and $|\phi\rangle$ are said to be orthogonal if their inner product satisfies

$$\langle \psi | \phi \rangle = 0.$$

Orthogonal states can be perfectly distinguished by a suitable measurement.

Conversely, two states are non-orthogonal if

$$\langle \psi | \phi \rangle \neq 0.$$

Non-orthogonal states cannot be perfectly distinguished by any measurement, even in principle. This intrinsic limitation forms the fundamental security mechanism of the B92 protocol.

*b) State Encoding by Mahendra:* In the B92 protocol, the sender (Mahendra) encodes classical bits using two non-orthogonal polarization states:

- Bit 0 → $|H\rangle$ (horizontal polarization, $0°$)
- Bit 1 → $|+\rangle$ (diagonal polarization

*c) Measurement Strategy of Emmanuel:* The receiver (Emmanuel) performs measurements using states that are orthogonal to Mahendra's signal states:

- $|V\rangle|$, orthogonal to $|H\rangle$
- $|-\rangle|$, orthogonal to $|+\rangle$

For each received qubit, Emmanuel randomly selects one of these measurement states.

*d) Detector Clicks and Bit Inference:* A detector click indicates that Emmanuel's measurement apparatus has detected a photon in the chosen measurement state.

- If Emmanuel measures $|V\rangle$ and obtains a click, Mahendra could not have sent $|H\rangle$. Therefore, Mahendra must have sent $|+\rangle$, corresponding to bit 1.
- If Emmanuel measures $|-\rangle$ and obtains a click, Mahendra could not have sent $|+\rangle$. Therefore, Mahendra must have sent $|H\rangle$, corresponding to bit 0.
- If no click occurs, the result is inconclusive, and the transmission is discarded.

*e) POVM Measurement in B92:* Emmanuel's measurement in the B92 protocol is best described by a Positive Operator Valued Measure (POVM) rather than a projective measurement. The POVM produces three possible outcomes:

- Conclusive result: bit 0
- Conclusive result: bit 1
- Inconclusive result

This three-outcome measurement allows Emmanuel to infer Mahendra's bit value with certainty in some cases, while safely discarding ambiguous events.

*f) Working of the B92 Protocol:*

1. State Preparation:
    Mahendra generates a random classical bit string and encodes each bit using the non-orthogonal states $|H\rangle|$ and $|+\rangle$.
2. Quantum Transmission:
    The encoded quantum states are transmitted to Emmanuel over a quantum channel.
3. Measurement:
    Emmanuel measures each received qubit using either $|V\rangle$ or $|-\rangle|$, chosen at random.
4. Announcement of Conclusive Events:
    Emmanuel publicly announces only the time slots in which he obtained a conclusive result, without revealing the measurement outcome.
5. Raw Key Generation:
    Mahendra identifies the bits corresponding to these time slots. These bits form the raw key.
6. Eavesdropping Detection:
    Mahendra and Emmanuel publicly compare a randomly selected subset of the raw key. If the observed error rate exceeds a predefined threshold, the protocol is aborted.

*g) Intercept Resend Attack and Error Analysis:* To illustrate the effect of eavesdropping, an intercept resend attack is considered. In this attack, an eavesdropper (Eve) intercepts the quantum states sent by Mahendra, performs a measurement, and resends the resulting collapsed state to Emmanuel.

Table III presents a representative sequence of transmissions involving Mahendra, Eve, and Emmanuel under such an attack. The table shows Mahendra's transmitted bits and states, Eve's measurement choices and outcomes, the states resent by Eve, Emmanuel's measurement choices, and the resulting inferred bits. Rows highlighted in red correspond to erroneous bit assignments caused by Eve's intervention.

TABLE III. *Bits sequence transmission*

| Row | Mahendra bits | Mahendra state | Eve | Eve Click | Eve Resend | Emmanuel Test | Emmanuel Click | Emmanuel bit |
|---|---|---|---|---|---|---|---|---|
| 1 | 0 | $|H\rangle$ | $|V\rangle$ | No | $|+\rangle$ (g) | $|-\rangle$ | Yes | 0 |
| 2 | 0 | $|H\rangle$ | $|-\rangle$ | Yes | $|H\rangle$ | $|V\rangle$ | No | - |
| 3 | 1 | $|+\rangle$ | $|V\rangle$ | Yes | $|+\rangle$ | $|V\rangle$ | Yes | 1 |
| 4 | 1 | $|+\rangle$ | $|-\rangle$ | No | $|H\rangle$ (g) | $|-\rangle$ | Yes | 0 |
| 5 | 0 | $|H\rangle$ | $|V\rangle$ | No | $|H\rangle$ (g) | $|-\rangle$ | Yes | 0 |
| 6 | 1 | $|+\rangle$ | $|-\rangle$ | No | $|+\rangle$ (g) | $|V\rangle$ | Yes | 1 |
| 7 | 0 | $|H\rangle$ | $|-\rangle$ | Yes | $|H\rangle$ | $|-\rangle$ | Yes | 0 |
| 8 | 1 | $|+\rangle$ | $|V\rangle$ | Yes | $|+\rangle$ | $|-\rangle$ | No | - |
| 9 | 0 | $|H\rangle$ | $|V\rangle$ | No | $|+\rangle$ (g) | $|V\rangle$ | Yes | 1 |
| 10 | 1 | $|+\rangle$ | $|-\rangle$ | No | $|H\rangle$ (g) | $|V\rangle$ | No | - |

*h) Quantum Bit Error Rate (QBER):* The Quantum Bit Error Rate (QBER) is defined as

$$QBER = \frac{N_{error}}{N_{conclusive}} \times 100\%$$

Where:

$N_{error}$ is the number of bits for which Mahendra's inferred bit differs from Emmanuel's original bit, and

$N_{conclusive}$ is the total number of conclusive measurement outcomes.

From Table III:

- Number of conclusive bits: $N_{conclusive} = 7$
- Number of erroneous bits: $N_{error} = 2$

$$QBER = \frac{2}{7} \times 100\% = 28.57\%$$

A QBER of this magnitude is significantly higher than expected in an ideal, eavesdropper-free scenario and clearly indicates the presence of an eavesdropper.

*i) Security of the B92 Protocol:* The security of the B92 protocol is based on these fundamental quantum principles

| Quantum Property | Role in Security |
|---|---|
| No-cloning theorem | Prevent Eve from copying quantum states |
| Measurement disturbance | Eve measurement introduce detectable error |
| Non-orthogonal states | Perfect discrimination is impossible |

Any attempt by an eavesdropper to gain information inevitably disturbs the quantum states and alters Mahendra's measurement statistics, enabling Emmanuel and Mahendra to detect the attack.

D. Performance Metrics

To evaluate protocol effectiveness, the following metrics are analyzed:

- Sifted Key Rate: The proportion of transmitted bits retained after basis comparison.

- Quantum Bit Error Rate (QBER): The ratio of incorrect bits to total compared bits, indicating channel reliability and potential eavesdropping.

- Security Robustness: The protocol's ability to detect interception and maintain key integrity under noisy conditions.

These metrics enable a consistent comparison of BB84, B92, and E91 protocols under simulated real-world conditions.

## IV. RESULTS AND DISCUSSION

This section presents and discusses the results obtained from the simulation of QKD protocols, namely BB84, B92, and E91. The evaluation focuses on key generation behavior, error characteristics, and protocol robustness under simulated realistic conditions, including basis mismatch and measurement uncertainty.

A. BB84 Protocol Simulation Results.

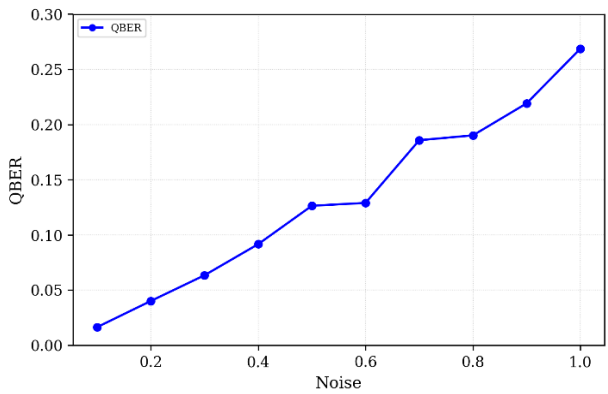

*Fig. 10. Relationship between Noise and QBER*

Figure 10 shows the relationship between noise and QBER for the BB84 protocol. The noise increases steadily as QBER rises, indicating a direct correlation between channel errors and noise levels. Higher QBER values result in increased noise, which degrades the performance and security of the protocol.

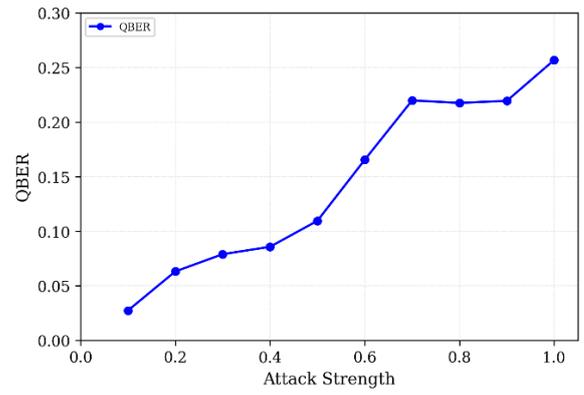

*Fig. 11. Relationship between Attack and QBER*

Figure 11 shows the variation of attack probability with respect to the Quantum Bit Error Rate (QBER) for the BB84 protocol. As QBER increases, the attack probability also increases, indicating a higher vulnerability of the quantum channel at elevated error rates.

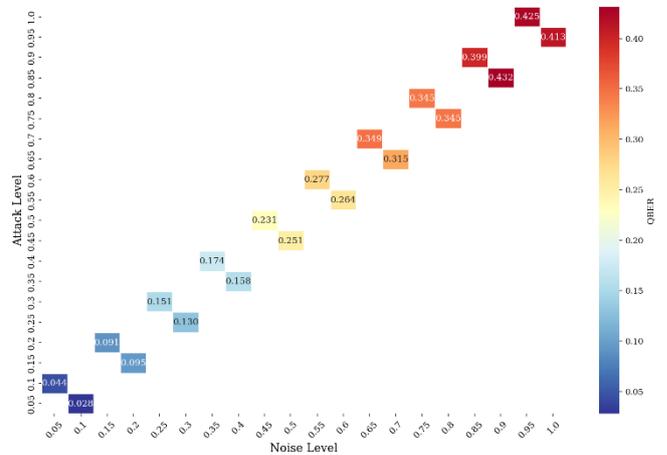

*Fig. 12. QBER heatmap for BB84 Protocol.*

Figure 12 shows a heat map of QBER as a function of noise level and attack level for the BB84 protocol. The color scale represents the magnitude of QBER, where blue indicates lower QBER values and red indicates higher QBER values. As both noise and attack levels increase, the color gradually shifts from blue to red, demonstrating a corresponding increase in QBER. This trend highlights the combined effect of channel noise and adversarial attacks on the performance and security of the BB84 protocol.

Risk Classification Based on QBER:

*Lowest Risk (0% ≤ QBER ≤ 4%):* When QBER is between 0% and 4%, the BB84 protocol is considered secure. The error rate is within the tolerable security threshold, meaning any disturbance from noise or eavesdropping is limited. Secure key generation is still reliable after error correction and privacy amplification.

*Mid Risk (5% ≤ QBER ≤ 11%):* When QBER falls between 5% and 11%, security becomes marginal. The error rate approaches or slightly exceeds the typical BB84 threshold, indicating significant disturbance. The protocol may still function, but the final key rate is reduced and security is less guaranteed.

*Highest Risk (QBER > 11%):* When QBER exceeds 11%, the protocol is considered insecure. The high error rate suggests strong eavesdropping or severe noise in the channel. At this level, secure key generation is no longer reliable, and the communication should be aborted

B. E91 Protocol Simulation Results.

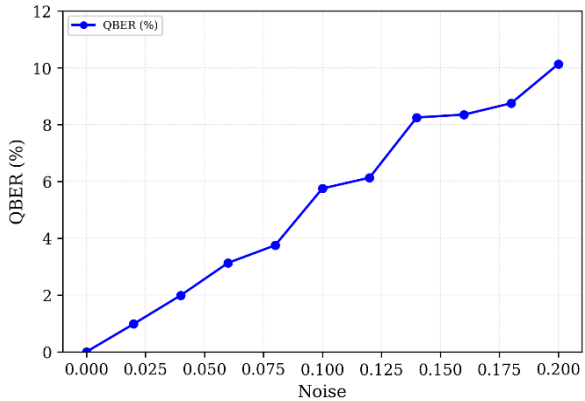

*Fig. 13. QBER as a function of channel noise in the E91 protocol with no eavesdropper.*

Figure 13 shows the variation of QBER as a function of channel noise in the absence of an eavesdropper. As the noise probability increases, the QBER exhibits a clear monotonic rise, indicating progressive degradation of key integrity due to environmental noise and imperfect quantum operations. This behavior confirms that channel noise directly impacts the reliability of the generated key, even when no adversarial attack is present.

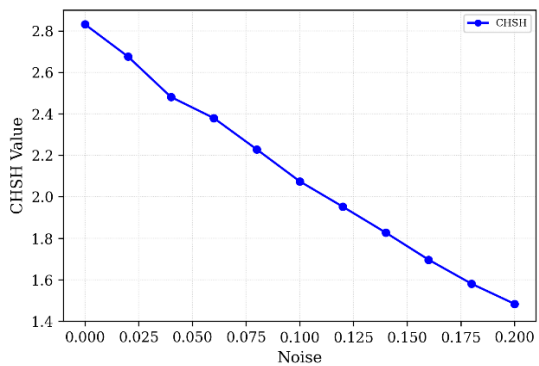

*Fig. 14. CHSH Bell parameter as a function of channel noise in the E91 protocol with no eavesdropper.*

Figure 14 illustrates the corresponding effect of channel noise on the CHSH Bell parameter. As observed, the CHSH value decreases steadily with increasing noise and eventually falls below the classical bound, indicating a loss of quantum nonlocality. This reduction demonstrates that excessive noise alone can destroy entanglement and invalidate the security guarantees of the E91 protocol.

Taken together, figures 13 and 14 demonstrate that channel noise adversely affects both key integrity and Bell inequality violation. While QBER captures the degradation of classical key correlations, the CHSH parameter reflects the loss of underlying quantum correlations. These results highlight the necessity of maintaining low noise quantum channels for secure operation of the E91 protocol.

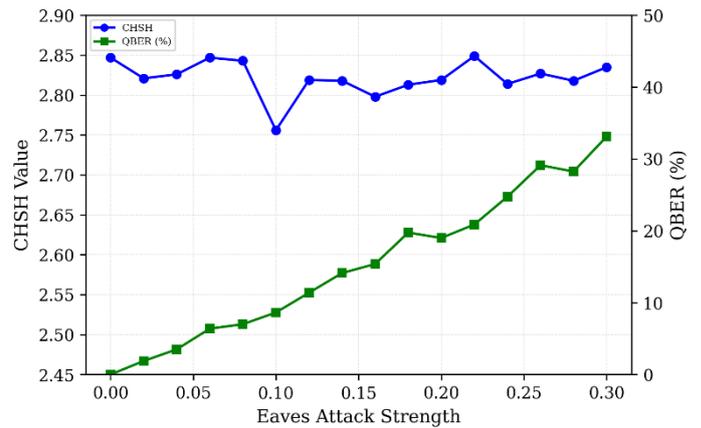

*Fig. 15. Effect of key only eave on CHSH parameter and QBER in the E91 Protocol*

As shown in figure 15, under a key only attack, the CHSH value remains close to the Tsirelson bound ($2\sqrt{2}$) across the full range of Eve probabilities, indicating that the entangled quantum correlations are preserved. However, figure 15 also shows a clear and monotonic increase in QBER, demonstrating that key-only attacks primarily degrade classical key integrity without affecting Bell violation.

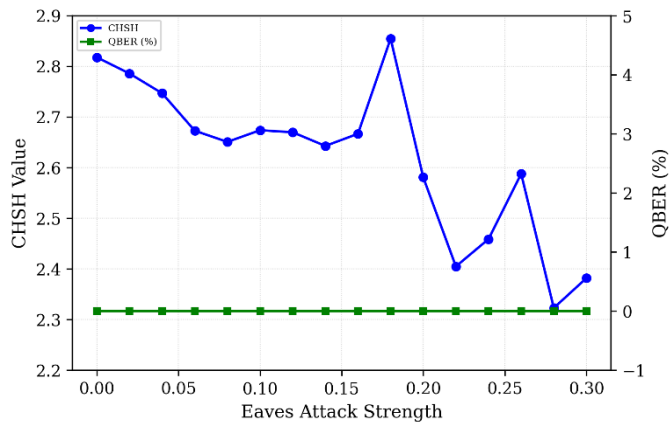

*Fig. 16. Effect of Bell only eavesdropping on the CHSH parameter in the E91 protocol.*

Figure 16 illustrates the impact of a Bell-only attack, where the CHSH value decreases steadily with increasing Eve probability and approaches the classical limit S=2, indicating a loss of quantum nonlocality. In this case, the QBER remains approximately zero, confirming that the key-generation process is not directly affected by Bell-only attacks.

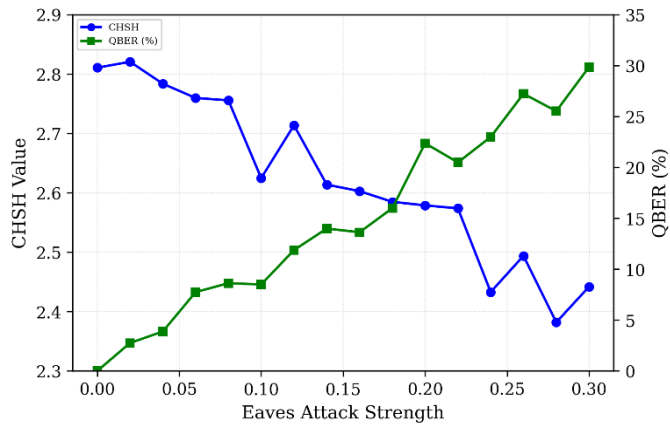

*Fig. 17. Effect of combined key and Bell eavesdropping on CHSH parameter and QBER in the E91 protocol.*

Figure 17 shows the effect of a combined key and Bell attack, the CHSH value decreases while the QBER increases significantly with Eve probability, indicating simultaneous degradation of quantum correlations and key consistency. This represents the most severe attack and leads to early detection and protocol abortion.

Together, Figures 15,16, and 17 demonstrate that the E91 protocol effectively detects different classes of eavesdropping through complementary monitoring of Bell inequality violation and quantum bit error rate, validating its robustness as a device independent quantum key distribution scheme

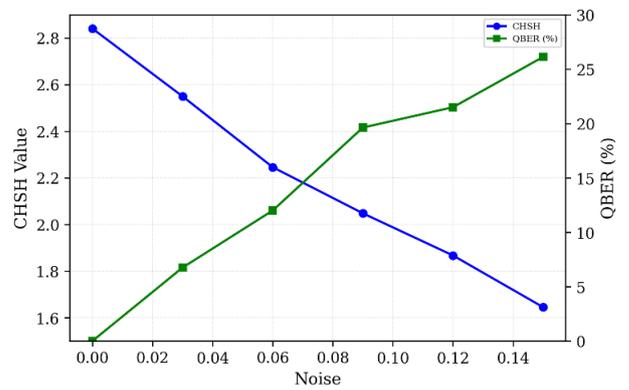

*Fig. 18. Increasing channel noise under combined Bell and key attacks results in simultaneous loss of Bell violation*

Figure 18 shows the combined effect of channel noise and eavesdropping on the CHSH Bell parameter and QBER in the E91 protocol under simultaneous key and Bell attacks. As the noise level increases, the CHSH value decreases steadily and falls below the classical bound, indicating a progressive loss of quantum nonlocality. At the same time, the QBER increases monotonically, reflecting significant degradation of key integrity due to both environmental noise and adversarial interference.

The simultaneous reduction of Bell violation and increase in QBER demonstrates that, under realistic channel conditions, the combined impact of noise and eavesdropping leads to rapid compromise of protocol security. This confirms that the E91 protocol effectively detects severe attack scenarios through complementary monitoring of Bell inequality violation and quantum bit error rate.

Risk Classification Based on CHSH and QBER:

*Lowest Risk (0% ≤ QBER ≤ 4% and CHSH > 2):* In the configuration of E91, the system is at the lowest risk when QBER is between 0% and 4% and CHSH remains clearly above 2. This indicates strong Bell inequality violation and minimal channel disturbance. Under this condition, entanglement is preserved and secure key distribution is reliable.

*Mid Risk (5% ≤ QBER ≤ 11% and CHSH approaching 2):* Mid risk occurs when QBER is between 5% and 11%, and CHSH begins to decrease toward the classical limit of 2. Although Bell violation may still exist, weakening entanglement and rising errors signal growing disturbance. Security is reduced and the system becomes less stable.

*Highest Risk (QBER > 11% and CHSH ≈ 2 or below):* Highest risk occurs when QBER exceeds 11% and CHSH approaches or falls near 2. This indicates significant channel disturbance and loss of strong quantum correlations. At this stage, secure key generation is unreliable and the protocol may need to be aborted.

## C. B92 Protocol Simulation Results.

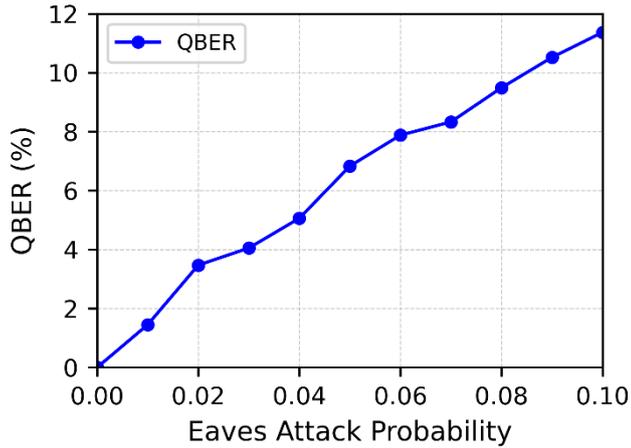

*Fig. 19. Quantum Bit Error Rate (QBER) vs. Eve Attack Probability in B92 Protocol (N = 20,000)*

Figure 19 illustrates the impact of increasing Eve's intercept resend attack probability on the Quantum Bit Error Rate (QBER). As the attack probability increases from 0 to 0.1, the QBER rises steadily, indicating that higher eavesdropping activity introduces more transmission errors. When no attack is present, the QBER remains near zero, confirming secure channel conditions.

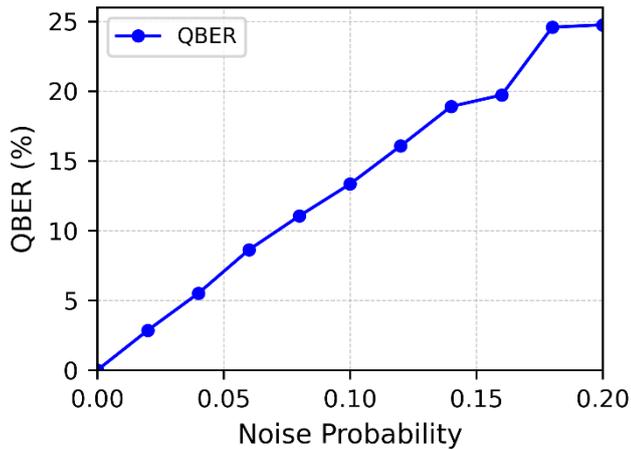

*Fig. 20. Quantum Bit Error Rate (QBER) vs. Channel Noise Probability in B92 Protocol (N = 20,000)*

Figure 20 shows the relationship between channel noise probability and the resulting Quantum Bit Error Rate (QBER) in the B92 protocol. As the noise probability increases from 0 to 0.2, the QBER increases nearly linearly, demonstrating that channel imperfections directly introduce transmission errors. When the channel noise is zero, the QBER remains negligible, confirming secure and error-free communication conditions.

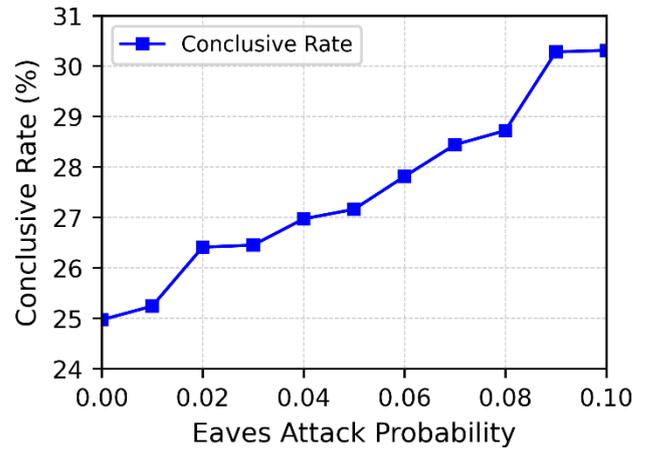

*Fig. 21. Conclusive Detection Rate vs. Eve Attack Probability in B92 Protocol (N = 20,000)*

Figure 21 illustrates the relationship between Eve's attack probability and the conclusive detection rate in the B92 protocol. The conclusive rate shows a slight increasing trend as the attack probability increases. This behavior indicates that intercept–resend attacks modify the quantum states reaching the receiver, thereby influencing the probability of obtaining conclusive measurement outcomes.

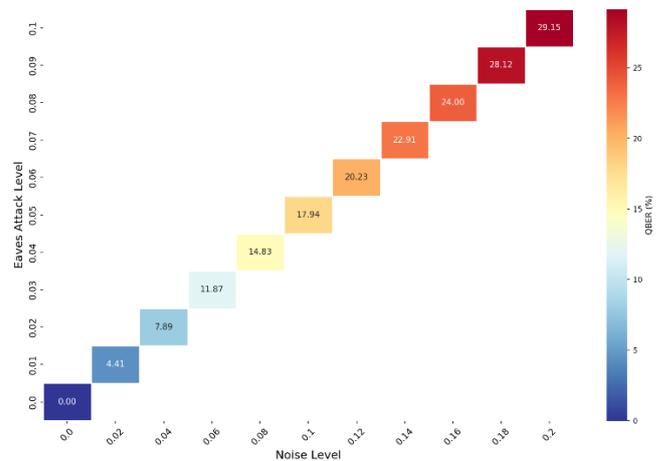

*Fig. 22. Heatmap Representation of QBER as a Function of Channel Noise and Eve Attack Probability (N = 20,000)*

The heatmap shows how QBER varies as both channel noise and Eve attack probability increase. The color scale represents the magnitude of QBER, where Blue shades indicate lower error rates and Red shades indicate higher error rates. As both parameters increase, the QBER rises steadily, demonstrating the combined impact of environmental noise and eavesdropping on system security.

Risk Classification Based on QBER:

*Lowest Risk (0% ≤ QBER ≤ 4%):* When the QBER is between 0% and 4%, the channel is considered secure. In this region, the observed error rate is within acceptable operational limits and can be attributed to minor channel imperfections or statistical fluctuations. Secure key generation can proceed without significant security concerns.

*Mid Risk (5% ≤ QBER ≤ 11%):* When the QBER lies between 5% and 11%, the channel is classified as compromised. This indicates noticeable disturbance that may arise from partial eavesdropping attempts or increased channel noise. Although key generation may still be possible, additional verification, error correction, and privacy amplification procedures are strongly recommended.

*Highest Risk (QBER > 11%):* When the QBER exceeds 11%, the system enters a critical risk region. Such a high error rate suggests significant interception activity or severe channel degradation. In this case, secure key establishment cannot be guaranteed, and the communication session should be terminated or reinitialized.

## V. CHALLENGES AND LIMITATIONS

Despite the strong theoretical security guarantees offered by Quantum Key Distribution (QKD), several challenges and limitations hinder its practical deployment. One of the primary constraints arises from the reliance on specialized quantum hardware. Real world QKD systems require dedicated physical channels such as optical fibers or satellite based links, along with precise photon sources and detectors. These requirements significantly increase implementation cost and complexity, making large-scale deployment difficult, particularly in academic and experimental environments. As a result, this study relies on simulation-based analysis, which, while flexible, cannot fully replicate all real hardware imperfections.

Another critical challenge is the susceptibility of quantum systems to noise and decoherence. Quantum states are highly sensitive to environmental disturbances such as thermal fluctuations, signal attenuation, and polarization drift. These effects introduce errors during transmission and measurement, increasing the Quantum Bit Error Rate (QBER). Accurately modeling these phenomena in simulations is complex, and simplified noise models may not capture all real world effects, potentially leading to optimistic performance estimates.

Scalability also remains a significant limitation. Most current QKD simulations and experimental implementations are restricted to point-to-point or small scale network configurations. Extending QKD to large scale or global communication networks introduces challenges related to key management, synchronization, and network coordination. Additionally, entanglement-based protocols such as E91 face further scalability issues due to the difficulty of generating and maintaining entangled states over long distances.

Finally, QKD research is inherently interdisciplinary, requiring expertise in quantum physics, cryptography, and computer science. This complexity increases the learning curve and poses challenges in system design, protocol implementation, and result interpretation. Addressing these limitations is essential for transitioning QKD from controlled experimental setups to practical, widely deployable quantum-secure communication systems.

## VI. CONCLUSION

Quantum Key Distribution represents a groundbreaking advancement in secure communication by leveraging principles of quantum mechanics to ensure provable security against eavesdropping, unlike classical cryptographic schemes vulnerable to quantum computing attacks. Through thorough simulation studies using frameworks such as IBM Qiskit, major QKD protocols like BB84, B92, and E91 have been modeled to analyze their performance under realistic conditions including quantum noise, decoherence, and potential interception attempts.

These simulations provide crucial insights into the operational efficiency, error rates, and robustness of different protocols, highlighting their strengths and practical limitations. While BB84 remains the foundational standard, continuous variable and entanglement based protocols offer promising avenues for integration with existing telecom infrastructure and complex quantum networks. The research underscores significant challenges due to hardware constraints, environmental noise, and scalability, which currently limit real world deployment but can be partially mitigated through advanced error correction, hybrid simulation approaches, and adaptive control techniques.

This work bridges the theoretical understanding and practical realization of QKD systems, laying a strong foundation for future quantum secure communication technologies integral to cybersecurity in the quantum era.

## VII. FUTURE WORK

The future research and development landscape of QKD simulation and deployment is rich with opportunities:

- **Enhanced Simulation Fidelity:** Incorporate more sophisticated modeling of quantum channel imperfections such as depolarization, timing jitter, and polarization drift to close the gap between simulations and physical experiments.

- **Scalable Network Simulation:** Extend simulation platforms to support large scale, multi node quantum networks, including satellite links, entangled node architectures, and measurement device independent QKD, enabling broader practical applicability.

- **Integration with Machine Learning:** Utilize advanced quantum machine learning and AI techniques to optimize protocol parameters, improve noise resilience, and detect sophisticated eavesdropping attacks dynamically.

- **Hybrid Quantum-Classical Systems:** Explore hybrid systems combining discrete variable and continuous variable QKD, and integrate quantum key distribution with classical cryptographic methods for layered security.

- **Hardware Software Co design:** Collaborate on developing integrated photonic chips and miniaturized hardware modules compatible with quantum simulators, improving experimental validation and field deployment readiness.

- **Standardization and Certification:** Develop universal composable security proofs, standardized benchmarking methods, and certification frameworks to accelerate commercial adoption and interoperability among diverse QKD deployments.

- **Educational Tools and Outreach:** Design portable simulation kits and educational platforms to facilitate broader understanding and adoption of quantum safe cryptography among researchers, students, and industry practitioners.